# Effect of an Optical Negative Index Thin Film on Optical Bistability

N. M. Litchinitser,[1] I. R. Gabitov,[2] A. I. Maimistov,[3] and V. M. Shalaev[4]

*1: Department of Electrical Engineering and Computer Science,*

*University of Michigan, 1301 Beal Avenue, Ann Arbor, Michigan 48109*

*2: Department of Mathematics, University of Arizona, 617 North Santa Rita Avenue,*

*Tucson, Arizona 85721, and Theoretical Division, Los Alamos National Laboratory,*

*Los Alamos, New Mexico 87545*

*3: Department of Solid State Physics, Moscow Engineering Physics Institute,*

*Kashirskoe sh. 31, Moscow, Russian Federation 115409*

*4: School of Electrical and Computer Engineering and Birck Nanotechnology Center,*

*Purdue University,*

*West Lafayette, Indiana 47907*

**Abstract:** We investigate nonlinear transmission in a layered structure consisting of a slab of positive index material (PIM) with Kerr-type nonlinearity and a sub-wavelength layer of linear negative index material (NIM) sandwiched between semi-infinite linear dielectrics. We find that a thin layer of NIM leads to significant changes in the hysteresis width when the nonlinear slab is illuminated at an angle near that of total internal reflection. Unidirectional diode-like transmission with enhanced operational range is demonstrated. These results may be useful for NIMs characterization and for designing novel NIMs based devices.

OCIS codes: (190.1450) Bistability, (310.6860) Thin films, optical properties, (160.4670) Optical materials

Recent experimental demonstrations of negative refractive index materials (NIMs) at optical frequencies [1-4] open a fundamentally new branch of modern optics and photonics and new possibilities for manipulating light waves. One of the most typical manifestations of negative refractive index is opposite directionality of the wave vector and the Poynting vector. As a result, many linear and nonlinear optical effects in NIMs differ from those in regular, positive index materials (PIMs) often in a very unusual way. Perhaps the two most striking effects are a reversed Snell's law in which a refracted light beam bends in the direction opposite to that in regular materials [5] and a "superlens" effect in which a plane-parallel slab of metamaterial can focus not only propagating waves as a regular lens does, but also the evanescent components [6]. Most of the theoretical studies of linear and nonlinear optical effects in NIMs to date focused on bulk NIM structures or periodic stacks of NIM and PIM layers [7-14]. However, currently these materials are only available in the form of a single very thin film with a thickness of about 150nm [1,2]. Obviously, effects like negative refraction or superlensing cannot be observed in such a thin film. The negative refraction property reveals itself in a phase shift (phase advance) that was measured interferometrically in the experiments reported in Refs. [1,2].

In this paper we investigate the effect of a very thin layer of NIM on the transmission properties of a bilayer consisting of a thin layer of NIMs and a nonlinear slab. We propose to utilize the phenomenon of optical bistability for NIMs characterization and novel device applications.

Most generally, optical bistability is a class of optical phenomena in which a system can exhibit two steady transmission states for the same input intensity [15].

Optical bistability has been predicted and experimentally realized in a various settings including a Fabry-Perot resonator filled with a nonlinear material [16], layered periodic structures [17] and a nonlinear slab [18-23]. A nonlinear film surrounded by a linear dielectric with high refractive index is known to exhibit bistability and more generally multistability when illuminated at an angle $\theta_{in}$, such that $\theta_{res} < \theta_{in} < \theta_{TIR}$, where $\theta_{res}$ is the incident angle corresponding to the resonant peak nearest to the angle of total internal reflection (TIR) $\theta_{TIR}$ in the linear transmission curve [20]. In this configuration, transmission in the linear regime is low. However, as the incident intensity increases, in the case of self-focusing Kerr nonlinearity, the nonlinear refractive index increases resulting in a shift of both $\theta_{TIR}$ and $\theta_{res}$ to larger values. Simultaneously, the transmission coefficient becomes multi-valued function of the input flux, leading to a bistable behavior.

The structure under consideration consists of a semi-infinite linear cladding (*cl1*), nonlinear optically thick layer – layer 1 (*NL*), linear NIM thin film – layer 2 (*NIM*), and a semi-infinite linear cladding (*cl2*) as shown in Fig. 1. We demonstrate that (i) an optically thin layer of NIM significantly modifies the bistable nonlinear transmission characteristics of this simple layered structure, and (ii) when the structure shown in Fig. 1 is illuminated leftwards versus rightwards, unidirectional transmission with increased operational range can be achieved.

The nonlinear layer is characterized by a dielectric permittivity

$$\varepsilon = \varepsilon^L + \varepsilon^{NL}\left(|E|^2\right), \tag{1}$$

where $\varepsilon^L$ is the linear part of the relative dielectric permittivity which generally can be complex, $\varepsilon^{NL}(|E|^2) = \varepsilon^L \varepsilon_0 c n_2 |E|^2$ is the nonlinear, intensity dependent part of the dielectric permittivity [19,20], $n_2$ is the nonlinear refractive index determining the nonlinear refractive index change given by $\Delta n_{NL} = \mu^L n_2 I$, where $\mu^L$ is the linear part of the relative magnetic permeability. Throughout this work it is assumed that the magnetic response is linear in all layers.

For TE polarization

$$E(x,z,t) = \frac{1}{2}[E(z)\exp[ik(\beta x - ct)] + c.c.], \quad (2)$$

where $k = \omega/c$ is the wave vector. In the nonlinear layer the following wave equation is solved [19]

$$\frac{d^2 E}{d\zeta^2} + p^2 E + \mu^L \varepsilon^{NL}(|E|^2) E = 0 \quad (3)$$

where $\zeta = kz$, and $p^2 = \varepsilon^L \mu^L - \beta^2$, $\beta = (\varepsilon_{cl1} \mu_{cl1})^{1/2} \sin\theta_{in}$. In the simple case of absorptionless media with Kerr nonlinearity, Eq. (3) has a well-known analytical solution. In the most general case, both the NIM layer and the nonlinear layer are lossy and the nonlinearity is not necessarily of pure Kerr type. Therefore, Eq. (3) has to be solved numerically. We use a numerical approach described in detail in Ref. [19]. In this approach, the electric field and its derivative specified at $z = d_1 + d_2$ are used to integrate Eq. (3) by a standard Runge-Kutta method within the layer 2. Calculated values of the field and its derivative are then used as new initial values to integrate within the layer 1 finally giving $E(0)$ and $dE/d\zeta(0)$.

As a first step, we consider the angular dependence of the transmission coefficient defined as $T = \frac{p_{cl2}\mu^L_{cl1}}{p_{cl1}\mu^L_{cl2}}\frac{|E_T|^2}{|E_{in}|^2}$, where $p^2_{cl1} = \varepsilon^L_{cl1}\mu^L_{cl1} - \beta^2$, $p^2_{cl2} = \varepsilon^L_{cl2}\mu^L_{cl2} - \beta^2$ for the three cases: (i) nonlinear optically thick slab with $\varepsilon^L_1 = 2.46$, $\mu^L_1 = 1$, $d_1 = 2.5\mu m$, and $n_2 = 2\times 10^{-9}\ m^2/W$ surrounded by a linear dielectric with a higher dielectric constant, (ii) the same slab as in case (i) but combined with a thin NIM layer characterized by $\varepsilon^L_2 = -2.46$, $\mu^L_2 = -1$, $d_2 = 150 nm$, and (iii) the same slab combined with a thin PIM layer characterized by $\varepsilon^L_2 = 2.46$, $\mu^L_2 = 1$, $d_2 = 150 nm$. In all cases, cladding layers are made of linear dielectric with $\varepsilon^L_{cl1} = \varepsilon^L_{cl2} = 3.087$, $\mu^L_{cl1} = \mu^L_{cl2} = 1$. Figure 2(a) shows the angular dependences of the linear transmission coefficient for cases (i)-(iii). The vertical dashed line corresponds to the TIR angle $\theta_{TIR} = 63.21°$, the vertical solid line corresponds to the incident angle $\theta_{in} = 63.1°$ used in all simulations in this paper. The corresponding transmission coefficient as a function of the input flux, defined as $S_{in} = \frac{\varepsilon_0 c p_{cl1}}{2\mu^L_{cl1}}|E_{in}|^2$, is shown in Fig. 2(b). Figure 2(a) shows that in the presence of the NIM (PIM) thin film, transmission maxima in the linear transmission spectrum shift to smaller(larger) angles with respect to the case of a single nonlinear layer. In general, transmission resonances are also affected by Fresnel reflections at the boundaries, however the choice of parameters in Fig. 2 guarantees perfect impedance match at the boundary between the nonlinear slab and the NIM(PIM) layer. The layered structure in Fig. 1 can be viewed as an analog of a resonator filled with intensity dependent material. In the linear regime, introduction of the NIM(PIM) layer is equivalent to decreasing(increasing) the length of the "resonator" ultimately leading to the transmission spectrum shift shown in Fig. 2. In the nonlinear regime, this analogy is valid if the thickness of the NIM(PIM) layer is small, so that an additional nonlinear phase shift that would be accumulated in the equivalent length of the resonator is negligible. Recall that we consider nonlinear transmission at incident angle near the TIR angle. As incident intensity continuously increases, the refractive index of the nonlinear slab also increases. Once the bistability threshold or switch-on point ("resonant condition") is reached, the output intensity switches from a low-transmission state to a high-transmission state. If the incident intensity is now slowly decreased, the system remains in the high-transmission state until the switch-off intensity is reached at which the system switches back to the low-transmission state. The input-output characteristic for such system forms a hysteresis loop. A linear NIM(PIM) thin film placed between the nonlinear slab and the second cladding (*cl2*) introduces an additional phase shift that affects the resonant condition. It is noteworthy that although the NIM(PIM) film is very thin, the effect of this phase shift on the nonlinear optical response of the entire structure turns out to be very significant. As discussed above, negative refraction reveals itself in a phase advance or a negative phase shift [1]. Therefore, in the case of a NIM thin layer, the "resonator" length decreases by $d_2$, implying that the intensity dependent nonlinear index change required for switching the transmission to the high-transmission state should increase. Therefore, we predict that the bistability threshold should increase in the case of the NIM, while the opposite effect should occur in the case of the PIM. Numerical results shown in Fig. 2(b) confirm this prediction.

Next, we study the effects of the NIM film parameters on the hysteresis curve and compare the results to those for the PIM case. Figure 3(a) shows the transmission coefficient versus input flux for fixed dielectric permittivity $\varepsilon_2^L = -0.5$ and varying magnetic permeability $\mu_2^L$ compared to those for PIM thin film with the same absolute values of $\varepsilon_2^L$ and $\mu_2^L$. The hysteresis width increases in the case of NIM film, as the absolute value of $\mu_2^L$ increases. In the PIM case, hysteresis width decreases and bistability almost disappears for $\mu_2^L = 4.5, \varepsilon_2^L = 0.5$. The bistability threshold increases as well in the NIM case, in agreement with the predictions of the simple resonator analogy based model discussed above. We find that the effect of the NIM thin film is less pronounced when the magnetic permeability is fixed, while the dielectric permittivity is varying (not shown here). We attribute the difference between the effect of $\varepsilon_2^L$ and $\mu_2^L$ to the fact that it is the magnetic permeability that affects the boundary conditions for TE wave. We expect that the effect of varying dielectric permittivity will be greatly enhanced for the TM wave as, in that case the boundary conditions are affected by the dielectric permittivity. Finally, we investigate the effect of varying both $\varepsilon_2^L$ and $\mu_2^L$ while the refractive index is fixed at $n = -1.5$ shown in Fig. 3(c) and at $n = 1.5$ in Fig. 3(d). Again, we find that the hysteresis curves' widths and switch-on points are significantly increased in the case of NIM while an opposite effect observed in the case of PIM.

Up to this point we assumed that the light always impinges the structure in Fig. 1 from the left. In what follows, we examine reciprocity in bilayer structure comprising a NIM film. Previously, it has been shown that bilayers comprised of two PIM materials, one of which possesses an intensity-dependent refractive index, exhibit bistability with spatially nonreciprocal character. For example, such structure is transparent for light incident from left to right while it is opaque for light traveling in the opposite direction [24]. Figure 4 shows hysteresis curves for the cases when the structure shown in Fig. 1 is illuminated rightwards versus leftwards for $\left|\varepsilon_2^L\right|=1$, $\left|\mu_2^L\right|=2$. Solid lines correspond to the NIM case when the structure is illuminated from left to right (1) and from right to left (2), dashed lines correspond to the PIM case when the structure is illuminated from left to right (3) and from right to left (4). We find that the range of intensities corresponding to unidirectional transmission is significantly increased in the case of a bilayer comprising thin film of NIM.

In summary, we studied numerically the nonlinear transmission properties of a layered structure consisting of a nonlinear slab and a thin layer of NIM. We found that even a very thin (subwavelength) film of NIM significantly modifies the nonlinear response of this structure. Owing to a very high sensitivity of the width and depth of the hysteresis to the changes of the material parameters of the NIM layer, optical bistability phenomena in these structures may be utilized for NIM characterization. In addition, we examined the nonlinear optical response as a function of directionality of the incident light and found a significantly increased (compared to the PIM case) range of input intensities corresponding to the unidirectional transmission. These results may be useful for characterization of NIMs as well as for novel device applications such as optical memory and optical diode.

The authors are grateful to Askhat Basharov and Vladimir Drachev for enlightening discussions. This work was supported in part by the NSF-NIRT Award # ECS-0210445, by the ARO grant W911NF-04-1-0350, by the ARO-MURI Award, by the NSF Award # DMS-050989, by the National Nuclear Security Administration of the U.S.

Figure captions:

Figure 1. Schematic diagram of the layered structure under investigation.

Figure 2. (a) The linear transmission coefficient versus the angle of incidence for a single layer 1 is shown by the solid line, the transmission coefficient for a combination of the layer 1 and a NIM thin film (layer 2) is shown by the dashed line, and the transmission coefficient for a combination of the layer 1 and a PIM thin film (layer 2) is shown by the dot-dashed line. (b) The transmission coefficient versus the input flux corresponding to the three configurations in Fig. 2(a).

Figure 3. Comparison of the transmission coefficient (in logarithmic scale) versus the input flux for the case of fixed dielectric permittivity in the NIM layer at $\varepsilon_2^L = -0.5$ (a) and in the PIM layer at $\varepsilon_2^L = 0.5$ (b) and varying magnetic permeability, and for the case of fixed refractive index in the NIM layer at $n = -1.5$ (c) and in the PIM layer at $n = 1.5$ (d) and varying both the magnetic permeability and the dielectric permittivity. Vertical straight lines indicate maximum bistability threshold in the PIM case.

Figure 4. The transmission coefficient versus the input flux for the layered structure shown in Fig. 1 when the light enters from the left (solid line 1) and when the light enters from the right (solid line 2). The dashed lines 3 and 4 correspond to the case of the PIM thin layer.

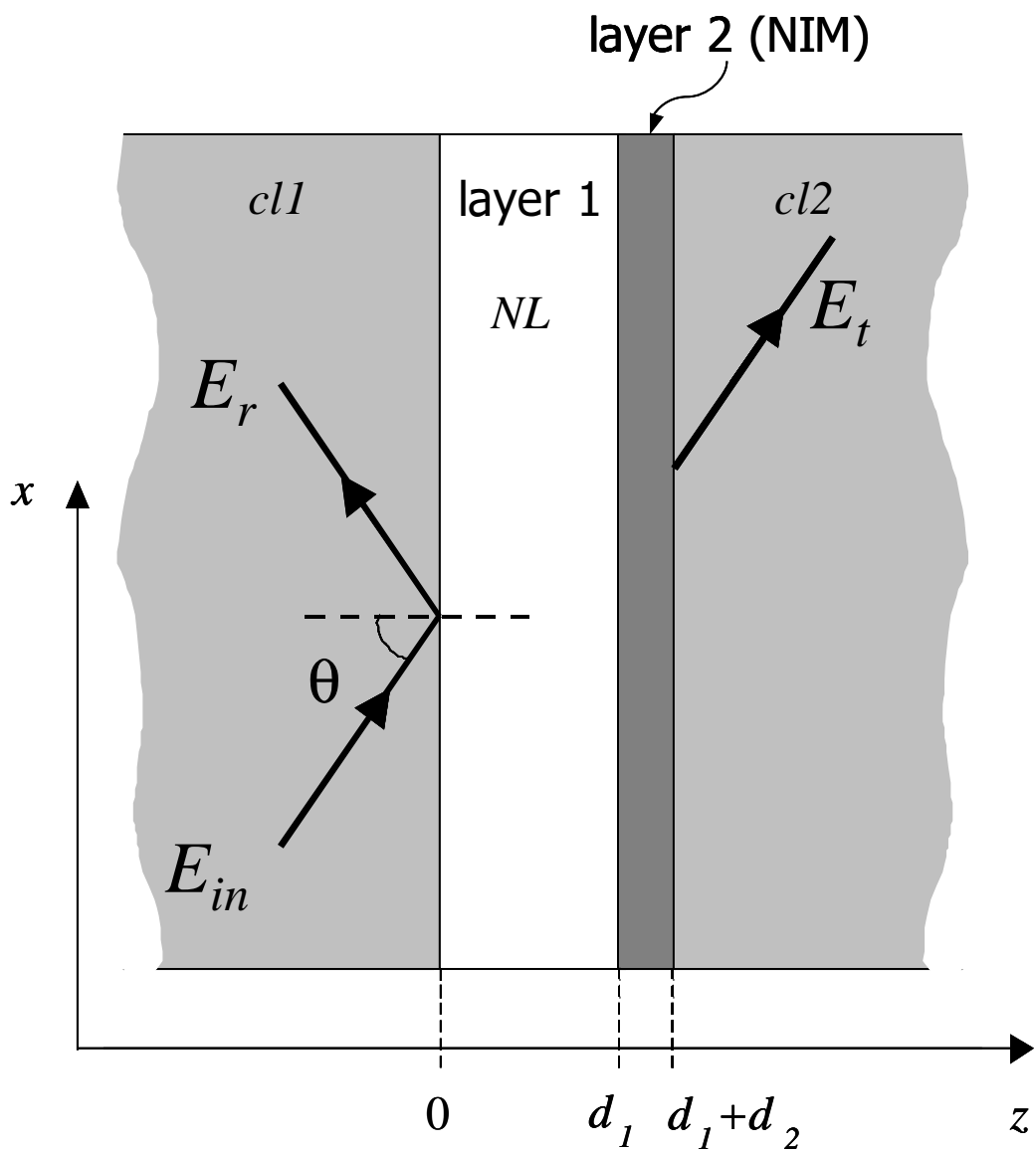

Fig. 1

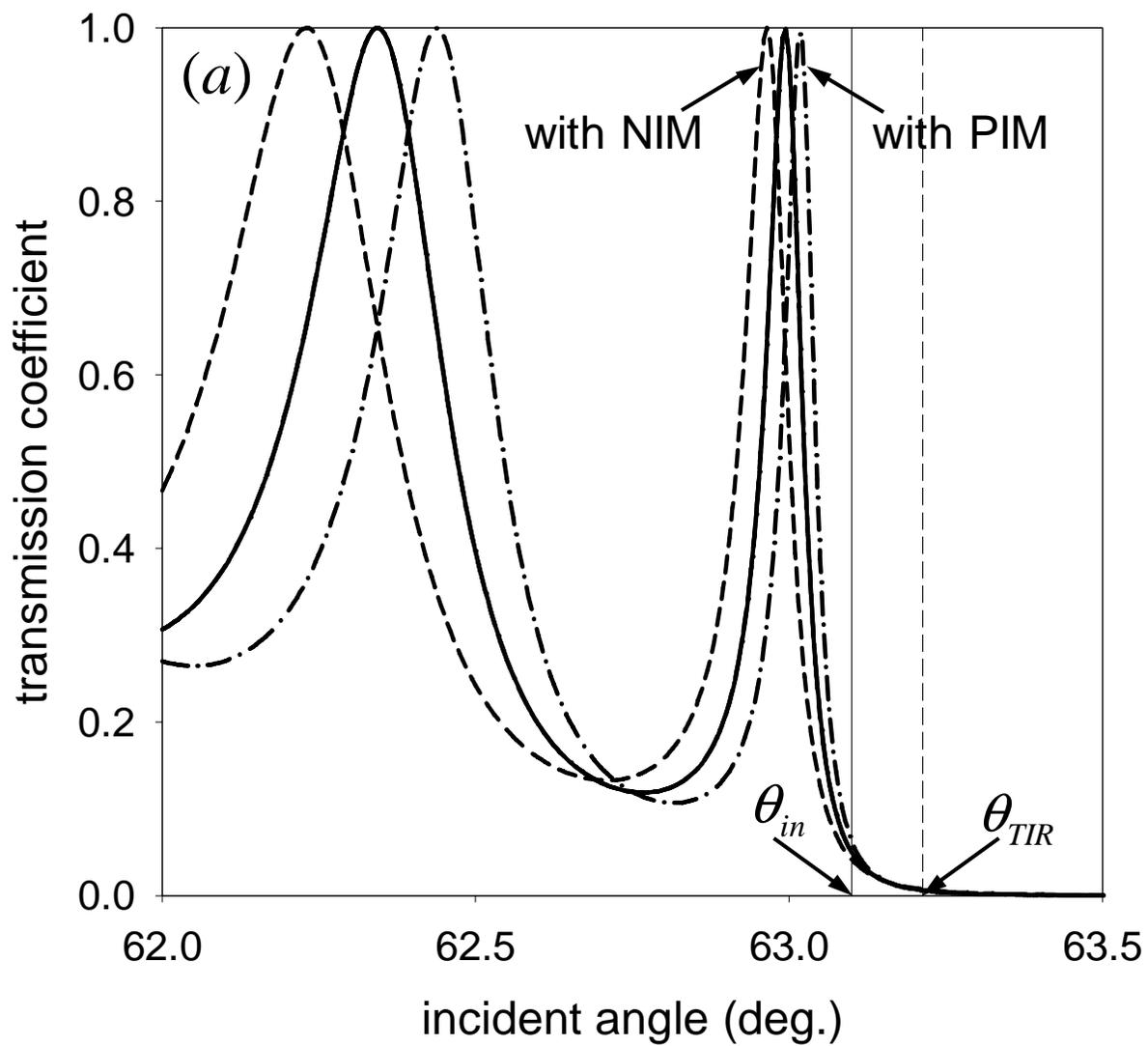

Fig. 2(a)

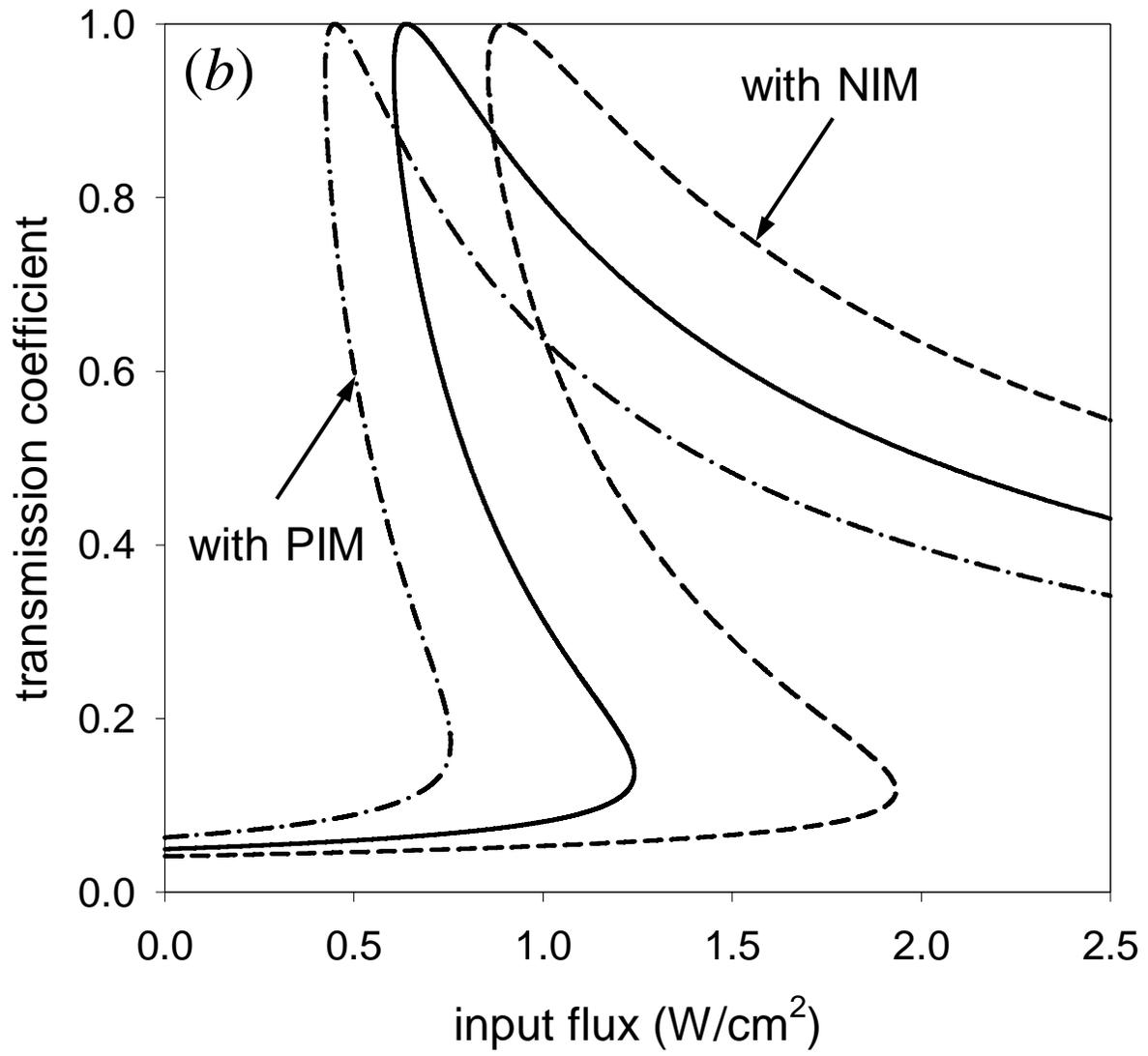

Fig. 2(b)

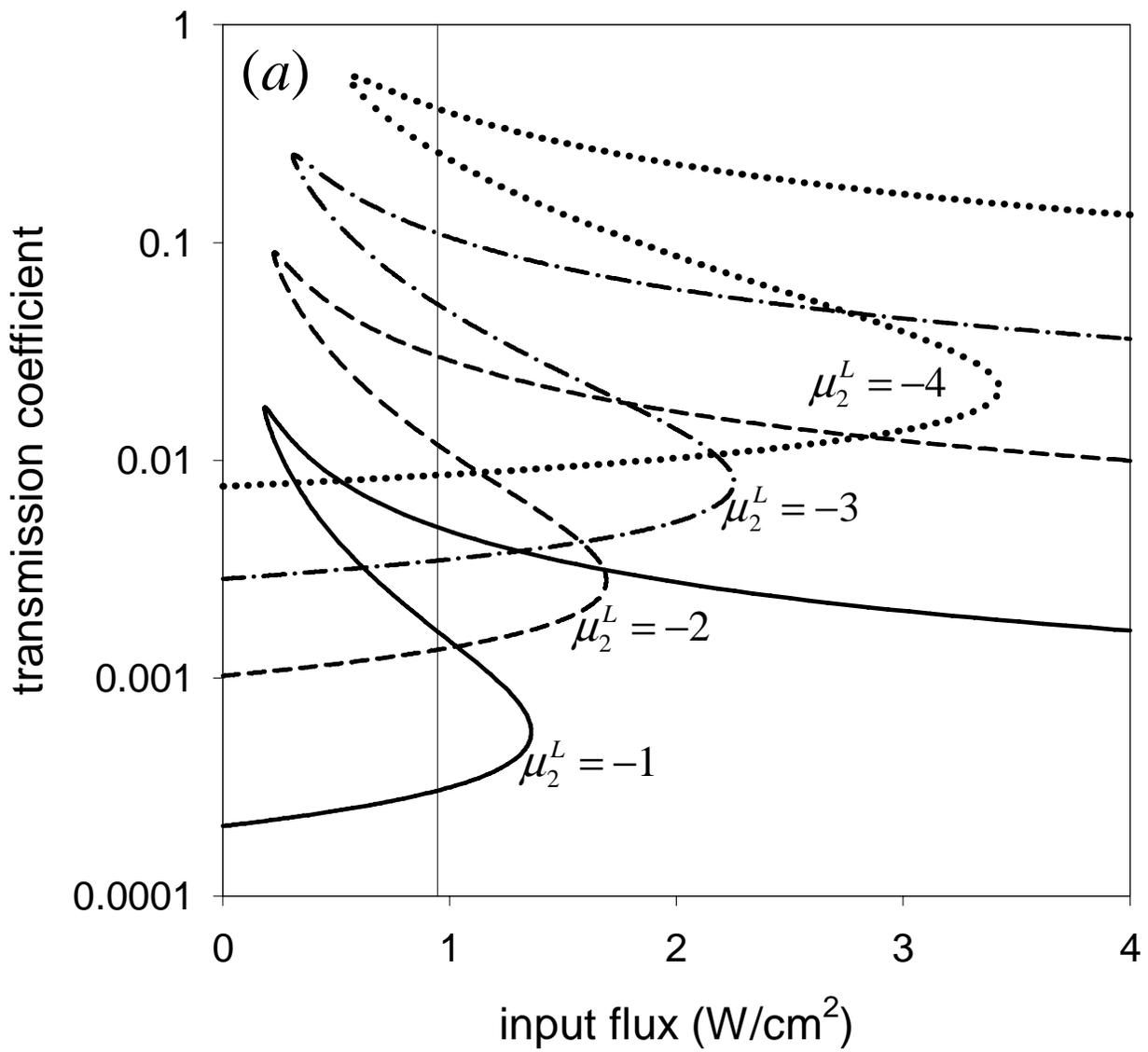

Fig. 3(a)

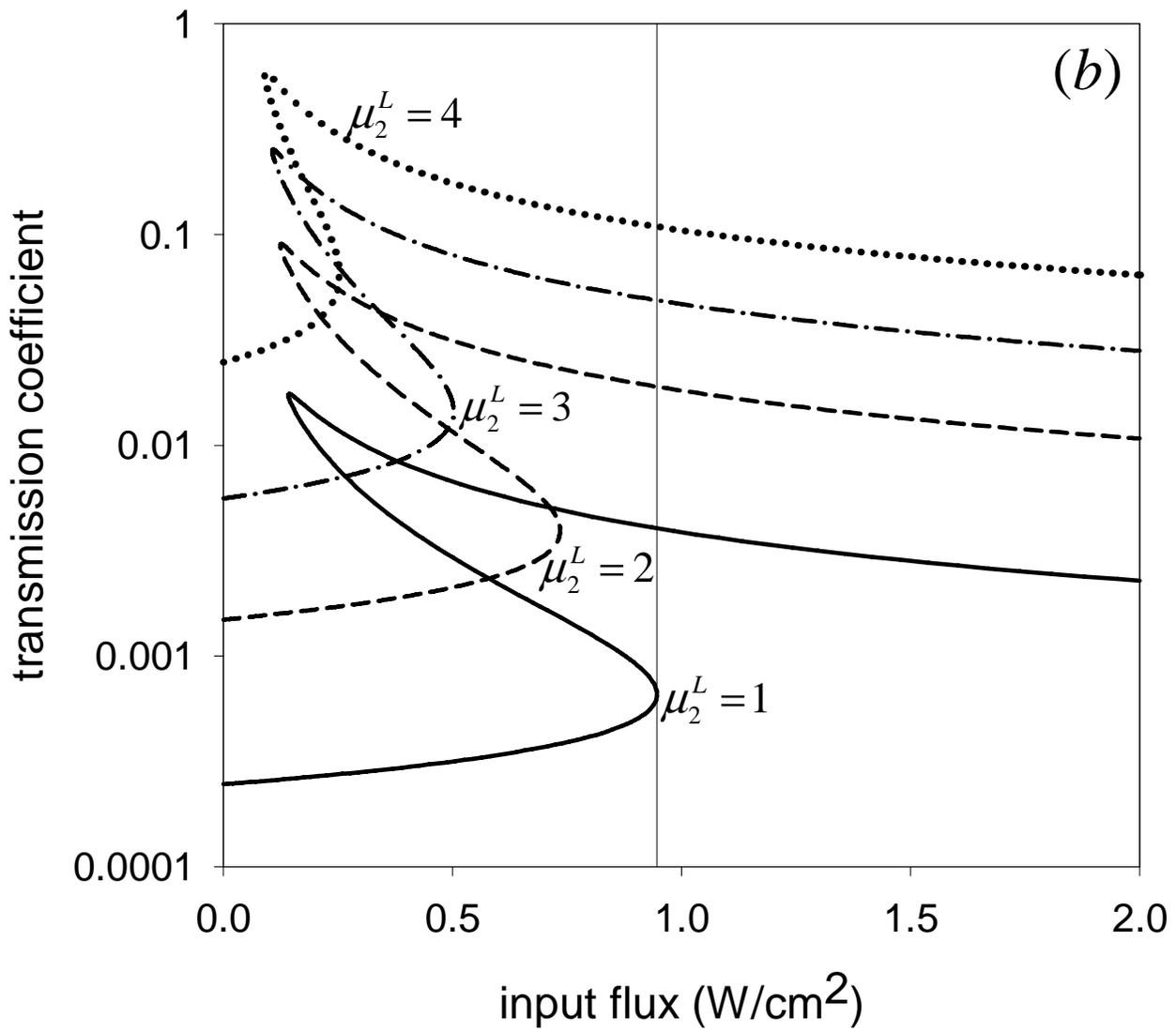

Fig. 3(b)

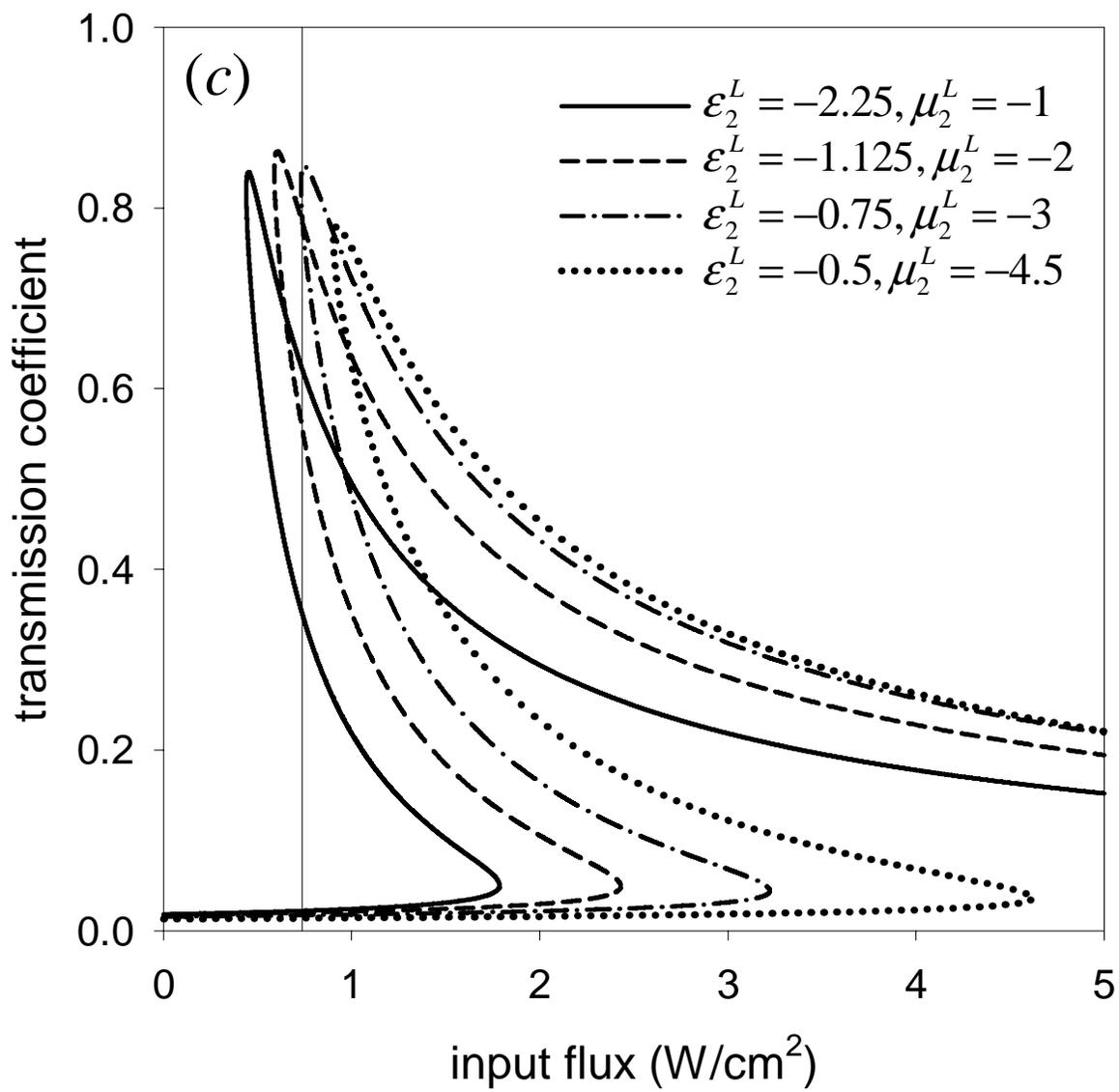

Fig. 3(c)

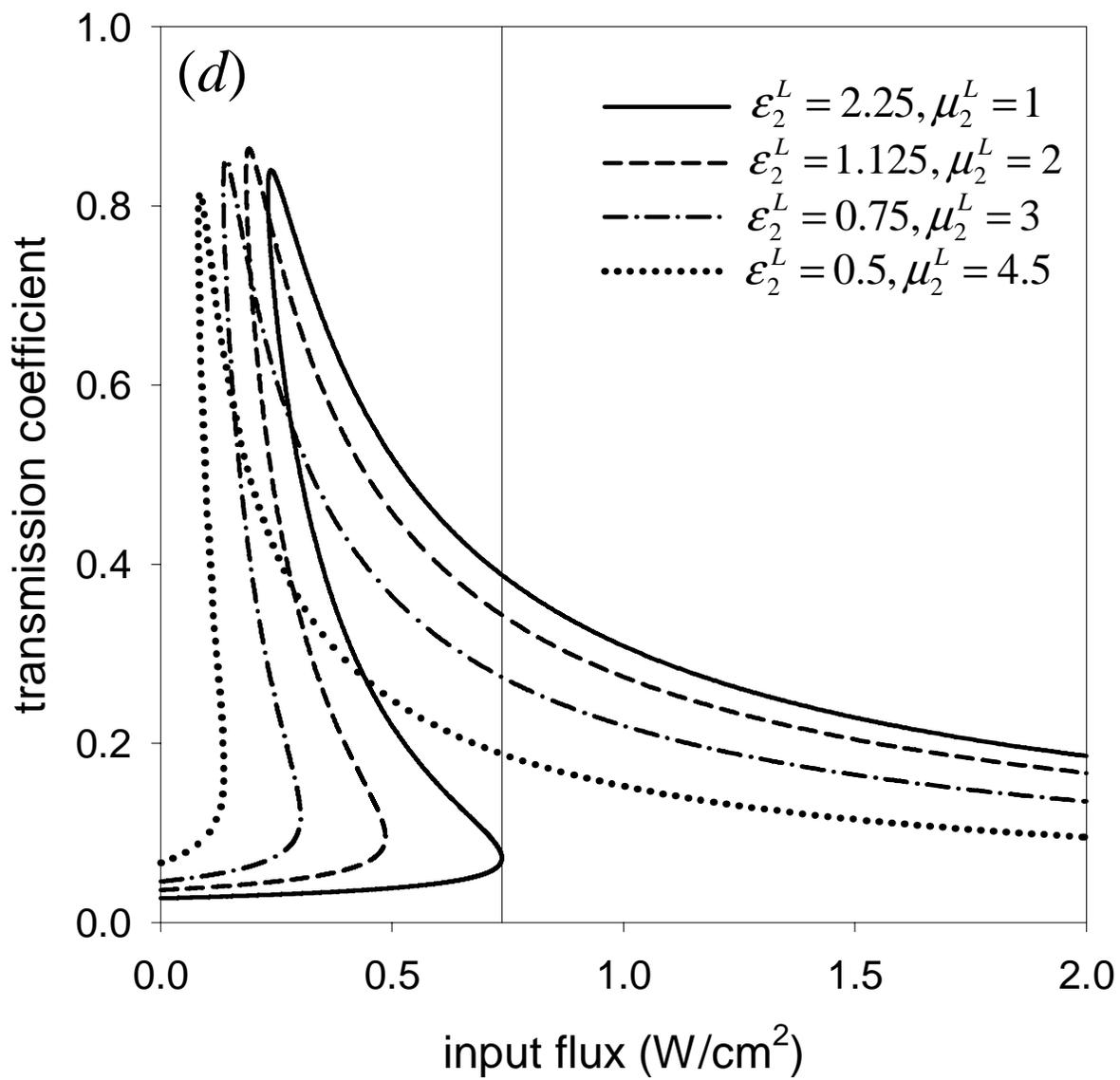

Fig. 3(d)

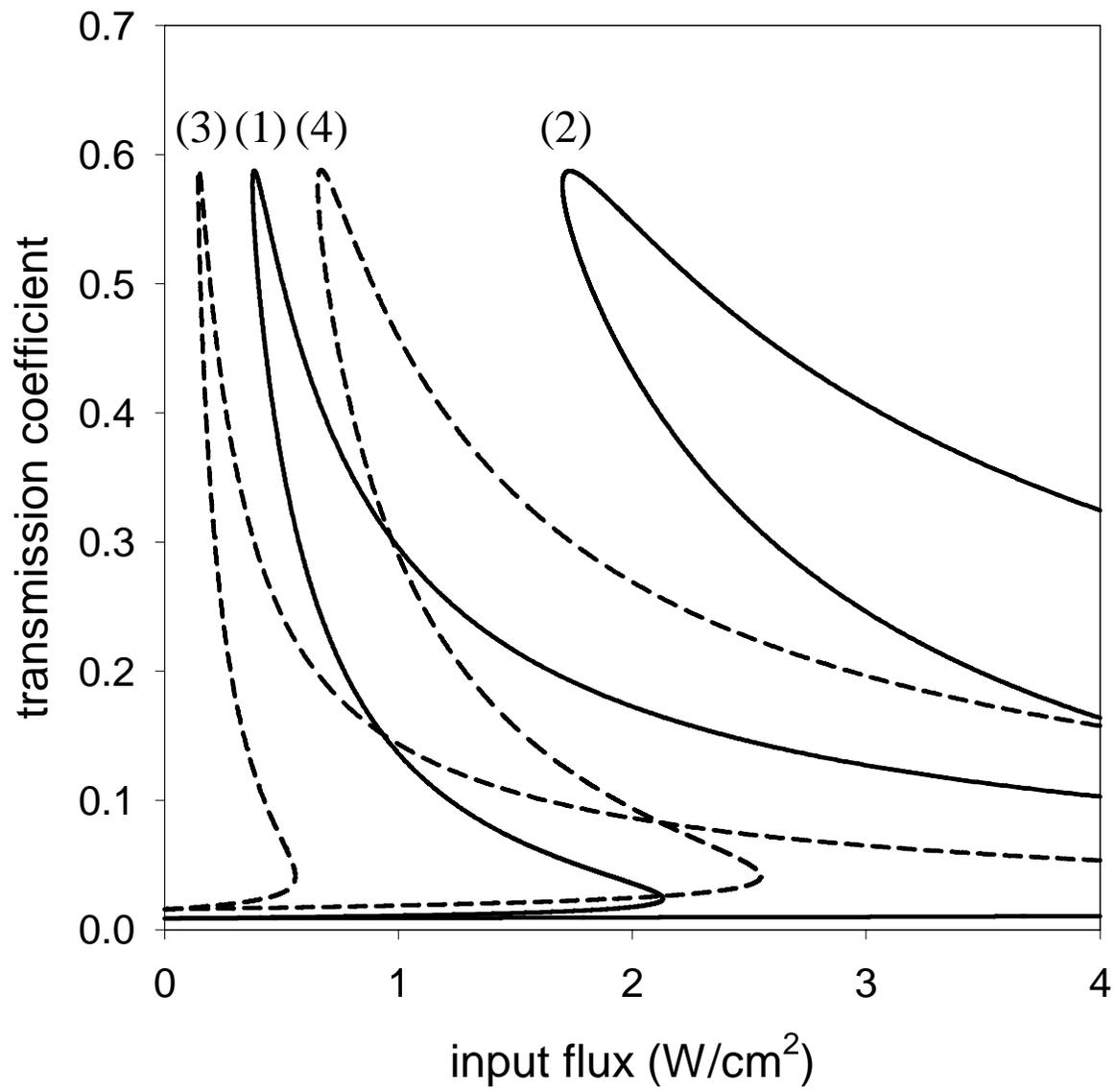

Fig. 4